\documentclass[sigconf]{acmart}
\usepackage{algorithm,algpseudocode}
\usepackage{graphicx}
\usepackage{array}
\usepackage{pgfplots}
\pgfplotsset{compat=1.18} 
\usepackage{amsmath}

\AtBeginDocument{%
  \providecommand\BibTeX{{%
    \normalfont B\kern-0.5em{\scshape i\kern-0.25em b}\kern-0.8em\TeX}}}

\setcopyright{acmcopyright}

\copyrightyear{2022}
\acmYear{2022}
\setcopyright{acmlicensed}\acmConference[GECCO '22]{Genetic and Evolutionary Computation Conference}{July 9--13, 2022}{Boston, MA, USA}
\acmBooktitle{Genetic and Evolutionary Computation Conference (GECCO '22), July 9--13, 2022, Boston, MA, USA}
\acmPrice{15.00}
\acmDOI{10.1145/3512290.3528693}
\acmISBN{978-1-4503-9237-2/22/07}

\begin{document}

\title{Adapting Novelty towards Generating Antigens for Antivirus systems }
\author{Ritwik Murali}
\authornote{Both authors contributed equally to this research.}
\email{m\_ritwik@cb.amrita.edu}
\orcid{0000-0002-1269-2257}

\authornotemark[1]

\author{C Shunmuga Velayutham}
\affiliation{%
  \institution{Dept. of Computer Science \& Engg.\\Amrita School of Engineering - Coimbatore,\\
Amrita Vishwa Vidyapeetham}
  \city{Coimbatore}
  \state{Tamil Nadu}
  \country{India}}
\email{cs\_velayutham@cb.amrita.edu}







\renewcommand{\shortauthors}{Murali,R and Velayutham, C S}

\begin{abstract}

It is well known that anti-malware scanners depend on malware signatures to identify malware. However, even minor modifications to malware code structure results in a change in the malware signature thus enabling the variant to evade detection by scanners.  Therefore, there exists the need for a proactively generated malware variant dataset to aid detection of such diverse variants by automated antivirus scanners. This paper proposes and demonstrates a generic assembly source code based framework that facilitates any evolutionary algorithm to generate diverse and potential variants of an input malware, while retaining its maliciousness, yet capable of evading antivirus scanners. Generic code transformation functions and a novelty search supported quality metric have been proposed as components of the framework to be used respectively as variation operators and fitness function, for evolutionary algorithms. The results demonstrate the effectiveness of the framework in generating diverse variants and the generated variants have been shown to evade  over 98\% of popular antivirus scanners. The malware variants evolved by the framework can serve as antigens to assist malware analysis engines to improve their malware detection algorithms.
\end{abstract}



\maketitle

\section{Introduction}
To the average user, a secure computer system is synonymous with the installation and usage of an automated antivirus software. These antivirus software are expected to detect and/or prevent malicious programs from affecting the end user's computing system. Most of the antivirus (AV) scanners identify malicious programs by examining new software/programs for predefined malicious patterns. These malicious patterns (called signatures) are identified previously by malware analysts and included in the antivirus scanner softwares database \cite{morley2001processing}, thus enabling automated detection.  While malicious software or malware have been prevalent since the early 1970s \footnote{Throughout this paper, the terms \textit{malware} and \textit{virus} are used interchangeably} \cite{szor2005art} and the use of up-to-date antivirus software has simplified the process of securing a computing system, it is a fact that the AV scanners still struggle to identify variants of malicious programs. This is because even minor modifications in the malware code result in a change in the known malicious pattern, thus enabling the malware to evade detection by the AV scanners \cite{malanov2012rapid}. The identification / detection of malware variants usually requires manual intervention and it is quite impractical for an analyst to predict and identify signatures of unknown malware variants \cite{murali2020malware}. The significance of the problem is further emphasized by the 2021 SonicWall Cyber Threat Report, where 589,313 new malware variants were identified by the SonicWall team in 2020 \cite{variants2020}. To put in perspective, this is a 57.35\% increase from 2019 where 153,909 malware variants were detected by the same company.\\
\par    

To counter this problem, anti-malware research has used two strategies with opposing perspectives. The first is a defensive strategy that applies Artificial Intelligence, Machine Learning \cite{zincir2021overview, yoo2021ai, jeon2020malware, bose2020explaining, maniarasusekar2020optimal, paul2017framework}, Data Mining \cite{coulter2020code,khalilian2018g3md,ye2017survey}, Evolutionary Algorithms \cite{wilkins2020cougar, manavi2019new, rafique2014evolutionary, mehdi2009imad}, Optimization Strategies \cite{macaskill2021scaling} and Knowledge Framework \cite{han2021aptmalinsight} based techniques to  detect and predict malware variants by identifying specific characteristics or features of the malware executable in order to detect and/or classify its variants. However, the performance (in terms of accuracy and false positives) of these techniques are dependant on the underlying data set used for training and classification and are severely affected by the shortage of such publicly available labelled data sets \cite{apruzzese2018effectiveness}. The second is a more aggressive / proactive strategy that involves creating malware variants using generative techniques, such as Generative Adversarial Networks (GANs), or meta-heuristic algorithms, like Evolutionary Algorithms (EAs), to reduce the real-world impact of the variants. While GANs exploit the non-linear structure of neural networks to generate complex adversarial examples capable of evading the target model \cite{aryal2021survey,li2020feature,tan2020enhancing,singh2019migan,hu2017generating}, EAs use biological evolution inspired strategies to generate malware capable of evading antivirus scanners \cite{divya2015real,murali2020preliminary}. In most cases, the proactive malware generation approach acts to augment the effectiveness of the defensive strategy as the malware variants generated serve as a valid database upon which the former can be trained or tested.    \\
\par

The use of EAs for malware generation is mainly focused on two approaches. The first involved the identification of various malware features with the EA being used to search for new variants with novel combinations of the identified features. The features were at the application level and the EA ensured that the malware variants generated have similar malicious characteristics. This was demonstrated by Noreen et al.\cite{noreen2009evolvable} who used a version of the email worm \textit{Bagle} to illustrate the application of evolutionary algorithms in malware generation. The worm was represented as a genome, which is essentially a collection of all the attack features such as date, port number, domain, email body, email subject, etc. The evolution involved searching for different combinations of attack features' values facilitated by mutation. The work also served as a framework for the application level approach and also used genetic algorithms as an evolutionary tool while showcasing a comparative study of different parent selection and crossover techniques that could be used for the malware, more specifically, computer worm generation. \\
\par
The second approach involves the use of the underlying assembly code of the malware as the platform for variation. Since all executables (irrespective of the source
programming language) can be disassembled to their assembly code structure, the second approach is gaining popularity. The explored approaches include using byte level modifications of the Windows Portable Executable (PE) files  \cite{castro2019aimed}, evolving entropy based polymorphic packers \cite{menendez2021getting}, identifying locations within valid executable to hide malicious code \cite{cani2014towards} and even a bottom up strategy of malware generation using Backus–Naur form (BNF) grammar to form the production rules to build code based on a designed grammar \cite{noreen2009using}. EAs have also been used to test the available AV solutions among mobile systems as well \cite{xue2017auditing,2016mystique,aydogan2015automatic} in an effort to confuse the AV systems to misclassify malware and/or evolve variants to test the AV systems themselves. \\
\par

While multiple algorithms are able to evolve an assortment of variants, the generation of a malware variant dataset, calls for creating a large number of malware variants possibly through code or feature modifications. In both cases, the modifications or transformations should also ensure and result in a diverse set of valid malware variants that serve as good representatives of the malware variants’ space.The focus so far, has broadly been on evolving variants that evade the antivirus scanners. Additionally, these strategies employ diverse malware representation and evolutionary operators as directed by the representation. Therefore, it is also increasingly challenging for the practitioner to validate the quality and innovations (in terms of complex code mutations) of the generated malware variants. \\
\par

Traditional evolutionary algorithms however are typically driven to converge to a fitness optimum. However, such a fitness strategy may not be apt when attempting to promote diversity and uniqueness in the resultant population. Novelty search is a divergent search algorithm that has the ability to promote evolvability \cite{lehman2011abandoning}. Evolvability allows the evolutionary algorithm to generate variability \cite{doncieux2020novelty}. This technique, inspired by natural evolution, rewards individuals that exhibit novel behaviours. The selection process for novel individuals therefore depends on the distance of the individual from its nearest neighbours in the behaviour space \cite{lehman2011improving}. Novelty search also suggests several strategies to manage the archive of past solutions. These include collecting the individuals whose novelty was above a threshold when first evaluated, the most novel individuals at each generation or even randomly chosen individuals if not none at all \cite{doncieux2019novelty}.\\
\par

In this work, we present an evolutionary algorithmic approach to evolve valid and potential (capable of evading effective anti-virus scanners) variants of a given malware. We consider the evolved (generated) malware variants as \textit{antigens} which can be presented to malware analysis engines to train and improve their malware detection algorithms thereby providing active acquired immunity to the end system against the existing numerous malware variants. We hypothesize that a novelty search based approach is capable of generating malware variants of greater diversity than a simple similarity based approach and verify the hypothesis with our Malware Antigens Generating Evolutionary algorithm (MAGE). The algorithm would serve as a foundation for an assembly source code based framework towards generating diverse, valid and potential variants (capable of evading antivirus scanners).  The modular design of the proposed framework facilitates extensibility in terms of malware representation, code transformation functions, quality indicator and the underlying evolutionary algorithm.

\section{Malware Antigen Generating Evolutionary algorithm (MAGE) as a Framework}

A framework that facilitates automated malware variant generation must handle two major challenges: (1) Ensure that the framework always results in a valid executable and the executable retains its maliciousness, and (2) The design of the framework should be flexible and modular such that the practitioner can induce application specific changes as per their requirement. In the context of an evolutionary algorithm ($E$), the automated malware variant evolution can be formally started as follows. To apply the evolutionary algorithm ($E$) on a given virus assembly code ($\zeta$) to generate variants ($\zeta'$) using assembly code transformation functions ($T$) for code mutations such that the maliciousness ($\Omega(\zeta')$) of the evolved variants is retained and the variant dataset generated by $E$ is diverse in nature.\\
\par
Therefore the proposed framework ($F$) includes the components $E, \zeta, \zeta', \Delta$ and $\xi $, where $T \in \{\psi_p, \tau_q, \sigma_r\}$ and $R$ are variation operators ($\Delta$), $\xi $ refers to the quality indicator or fitness function and $\psi,\tau$ and $\sigma$ are the control flow modifications, data transformations and code layout changes respectively. This implies that any $E$ can operate on $\zeta$ and employ $\Delta = \{T,R\}$ as variation operators and  $\xi $ to evaluate the candidates to realize the automated generation of valid variants $\zeta'$ of $\zeta$. In the perspective of an evolutionary algorithm, the framework models the different representations possible in an assembly language environment and discusses a few of the variation operators ($T, R \in \Delta$) and uses a modified version of novelty as the fitness function ($\xi$). The general structure of a typical population based meta-heuristic algorithm is given in Algorithm \ref{Algo1:StructureOfEA}. Each module in the framework is flexible enough to be customized as per the practitioners requirement as can be observed during the description of the Malware Antigens Generating Evolutionary algorithm (MAGE) in the next section.

\begin{algorithm}
  \caption{Structure of an evolutionary algorithm}
  \label{Algo1:StructureOfEA}
  \begin{algorithmic}[1]
    \Procedure{Evolutionary Algo}{$ $}
      \State INITIALIZE a population using random individual solutions
      \State EVALUATE every individual in current population
      \While{NOT TERMINATION CONDITION}
        \State \texttt{SELECT parents using any selection strategy}
        \State \texttt{RECOMBINE the parent pairs} \Comment{i.e. perform crossover}
        \State \texttt{MUTATE the resultant child} \Comment{i.e. the new individual}
        \State \texttt{EVALUATE the new individual}
        \State \texttt{SELECT individuals for the next generation}
      \EndWhile
    \EndProcedure
  \end{algorithmic}
\end{algorithm}

\subsection{Representation}
\begin{figure}
    \centering
    \includegraphics[width=1.0in,height =1.8in]{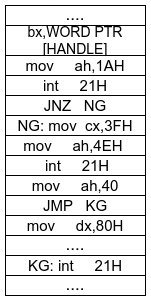}
    \caption{Linear representation of virus code.}
    \label{fig:LinearRepresentation}
\end{figure}

\begin{figure}
    \centering
    \includegraphics[width=2.5in,height =2.3in]{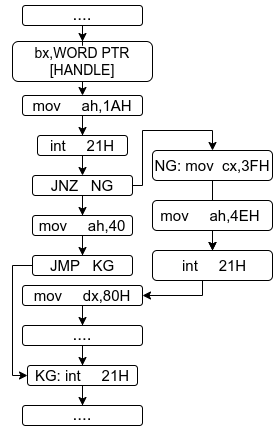}
    \caption{Graph representation of virus code.}
    \label{fig:GraphRepresentation}
\end{figure}

From the perspective of the evolutionary algorithm, considering the case of the assembly language code structure, there are a number of ways the source code can be represented. Here each individual in the population is the entire assembly code (program) of the chosen malware (or even a variant). The representation is quite flexible and allows the individual to be modelled either in a linear fashion (figure \ref{fig:LinearRepresentation}) or as a graph (figure \ref{fig:GraphRepresentation}). Both the framework and the evolutionary algorithm proposed in this work adopt the linear representation of the malware code for the simulation experiments. 

\subsection{Quality indicator}
The quality indicator $\xi$ (fitness function in EA) of the framework quantifies the extent of the transformation introduced in the source assembly code. In the context of malware variant generation, this work attempts to explore a variation of novelty search as a strategy to yield divergent code structures. Considering each individual as a vector of assembly code statements (as the linear representation dictates), the quality indicator ($\xi$) is calculated as the euclidean distance between each individual in the population and the mean vector ($\Bar{S}$) of all intra-population similarity vectors in the population as shown in equation \ref{qualityIndicator}. Here P is the population size.   

\begin{equation}
\label{qualityIndicator}
    \xi = \sqrt{\sum_{i=1}^{P} (\Bar{S},\overrightarrow{SI_i})^2} 
\end{equation}


The similarity vector $\overrightarrow SI_i$ for the $i^{th}$ individual within the population P is $\overrightarrow{SI_i} = J(p_1, p_i), J(p_2 , p_i ), ..., J(p_{NP} , p_i ), J(\zeta, p_i )$  where $p_1, ..., p_{PG}$ is the rest of the population comprising all chromosomes except $p_i$ , $\zeta$ is the source virus code and $J(p_j, p_i )$ is the Jaccard similarity index between a $j^{th}$ and $i^{th}$ chromosome as shown in equation \ref{eq:jaccardFormul}.  
\begin{equation}
    \label{eq:jaccardFormul}
    J(h_j, h_i )= \frac{|h_j \cap h_i|}{|h_j \cup h_i|} =  \frac{h_j \cap h_i}{|h_j| + |h_i| - h_j \cap h_i}
\end{equation}
\\
\par

The Jaccard similarity index was chosen after preliminary experiments revealed that this was the most stringent metric for generating divergent malware. Nevertheless, it is worth reiterating that the choice(s) proposed for each key component of the framework is merely suggestive and a practitioner, by virtue of the modular nature of the framework, can employ their choice(s) for each key component.

\subsection{Variation Operations}
Given the representation, the variation operations $\Delta$ are defined as mutation and crossover operations. For the chosen assembly language representation, the framework proposes code transformation operators as both mutation ($T$) and crossover ($R$) operations.\\
\par

\subsubsection{Mutation: \\}
In the case of malware, the framework exploits the assembly level constructs commonly used by malware authors for evading antivirus scanners \cite{hosseinzadeh2018diversification}, to model the transformation operators ($T$). Therefore the transformation operators can be considered as a subset of the code evasion operators namely modifying control flow ($\psi$), Transforming Data ($\tau$)) and Changing Code Layout ($\sigma$) (i.e. $T \subseteq \{\psi,\tau,\sigma\}$) and this includes the use of opaque predicates ($T_{OP}$), bogus insertions ($T_{BI}$), branching functions ($T_{BF}$), instruction transformation ($T_{IT}$), code block reordering ($T_{RB}$), variable substitution ($T_{VS}$) and code block substitution ($T_{CBS}$). Each of the transformation operators are limited by a set of constraints ($C$) that define the allowed structure resulting from transformations thus limiting the potentially infinite productions and ensuring valid code structures.\\
\par
Based on generic transformation functions ($T_{OP},T_{BI},T_{BF},T_{IT}$, $T_{RB}, T_{VS}$ and $T_{CBS}$), five transformation function instances namely Fake Instruction ($T_{FI}$), Forced JMP ($T_{FJ}$), Untouchable Block ($T_{UB}$), Conditional Zero JMP ($T_{CZJ}$) and Conditional Non Zero JMP ($T_{CNZJ}$), have been defined and proposed. It is worth mentioning that every assembly language program code has a standard prologue (header), a program body and epilogue (footer). Therefore, all applications of code transformations are constrained by a common rule ($C_{com}$) which states that ``Every code transformation must be applied in the program body of the input assembly code''. It should be noted that only the common constraint $c_{com}$ is applicable in the case of \textit{Fake instructions} ( $T_{FI}$), such as `NOP', which can be inserted anywhere within the program body as it does not affect the code functionality in any manner. $C_{com}$ here ensures that a valid executable is created post application of the transformation. However, while multiple \textit{NOP} instructions can be inserted in any location within the program body, excessive insertion of this instruction would result in the code bloating thus increasing the chances of generating invalid executables.\\
\par

\begin{figure}[htb]
    \centering
     \scalebox{0.77}{
    \includegraphics[height=12cm, width=2.0in]{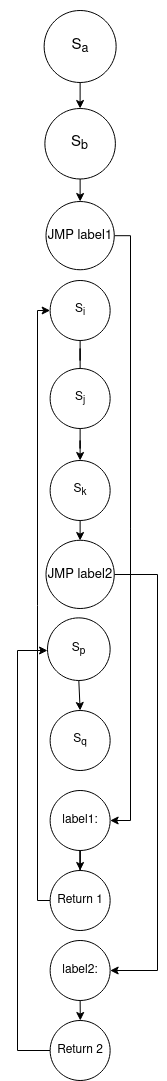}
    }
    \caption{Forced JMP Transformation}
    \label{fig:forcedJMP}
\end{figure}

``JMP'' is an assembly language instruction that alters the control flow of the assembly code and can be inserted anywhere within a contiguous code block. This is a logical choice as insertion of a JMP statement beyond the boundaries of any function is meaningless since the command will never be called and would act as dead code. The JMP instruction is always paired with a \textit{label} and this acts similar to a function call and in the framework $F$, must be defined beyond any continuous block of assembly code. This is to ensure that the label definition does not interrupt or affect any existing function blocks or contiguous code structures. The label definition should end with a reference/return to the line/command immediately after the JMP call so as to ensure continuity in the control flow of the code. This also ensures that  nesting of such a transformation results in complex but linear code flow as shown in figure \ref{fig:forcedJMP}. Based on the above understanding of the JMP statement, the \textit{Forced JMP} ($T_{FJ}$) transformation is a logical control flow modification choice for code mutations  and it is defined as $T_{FJ}=(C( JMP, \$label, s_1 , \$label:, s_1 ))$ where ``label'' refers to the logical position of the control flow transfer and $s_i$ stands for each subsequent assembly language instruction. The flexibility of the JMP statement can also be exploited to create the \textit{Untouchable Block} transformation, formally defined as $T_{UB}=(C( JMP, \$label, s_k^+ , \$label: ))$. Here, immediately after the JMP statement multiple assembly language statements ($s_k^+$) may be inserted. However, since none of these instructions will ever be executed, all of them act as dead code. The lines of code and structure of the code is changed here while retaining the control flow. \textit{Conditional JMP} ($T_{CZJ}$ and $T_{CNZJ}$) transformations check the status of the zero flag at that specific instance of code execution before executing a JMP instruction. This further increases the number of code level variations that can be evolved by the evolutionary algorithm. 

There are also certain transformation functions that are classified under $\psi, \tau$ and $\sigma$, commonly used in higher level programming languages. These include operations such as loop unrolling, class transformations, array transformations, etc., which are not feasible to apply in the context of assembly language transformations because the format is not supported in assembly languages and/or they do not always result in a valid executable post application on code. Therefore such transformations have not been considered in the framework proposed. The chosen transformations ensure that the transformed code is always a valid executable and thus form potential candidates for mutation operations $M \in \Delta$.

\subsubsection{Crossover: \\}
The crossover or recombination operation ($RE \in \Delta$) is another integral variation operation of an evolutionary algorithm. In the context of the code transformation functions described above, the recombination operation can be realized as \textit{code block transformation} $(T_{CBI})$ and can be defined as follows. If P is the population of all individuals in the solution space of a single generation then  $p$ and $\hat{p}$ in the population represent two candidate parents for the crossover operation. In the context of the proposed assembly code based framework, $p=\{S^{+}, LOP^{+}, COP^{+}\}$ and $\hat{p}=\{\hat{S}^{+}, \hat{LOP}^{+}, \hat{COP}^{+}\}$. Then $\langle O,\hat{O}\rangle :=RE(p,\hat{p})  :=  T_{CBI}\{(S^+,LOP^+,COP^+),(\hat{S}^+,\hat{LOP}^+,\hat{COP}^+), C_i\} $ where $O$, $\hat{O}$ are the offsprings resulting from recombination operation, $LOP$ \& $COP$ represent the loop \& conditional statements in the assembly language construct, and $T_{CBI}$ represents code block interchange transformation function. Table \ref{Table:TransformationInstances} shows a summary of the instances of the variation operators.\\
\par

\par
The code block interchange transformation function $T_{CBI}$ interchanges blocks of code that might contain statements as well as loop and conditional operands. In the absence of a constraint, $T_{CBI}$ has the potential to yield invalid executables as offsprings and hence can be very disruptive. Consequently, the choice of code block location, (henceforth called pivot point) is very crucial to ensure that code block interchange does not interrupt the sequential execution of the resultant assembly code. The constraint then requires the interchange to be applied between code blocks above the pivot point and similarly swap code blocks below the pivot point as well. This constraint reduces the possibility of a label defined by any transformation function being lost during interchange operation else an undefined label would result in an invalid executable. \\
\par
\begin{figure}[htbp]
    \centering
    \includegraphics[scale=0.36]{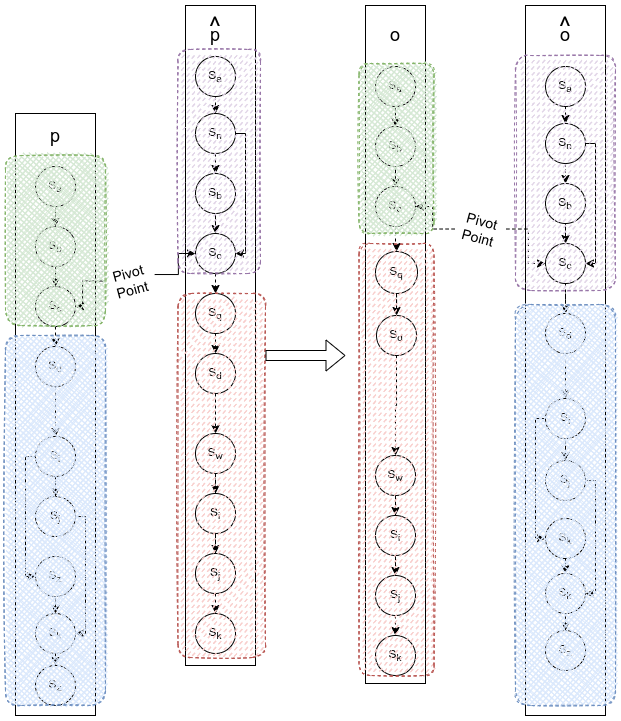}
    \caption{Code block interchange transformation}
    \label{fig:pivotPoint}
\end{figure}
\par

\begin{table}[htbp]
\caption{Summary of variation operator instances}
\label{Table:TransformationInstances}
\scalebox{0.79}{
\begin{tabular}{|l|l|l|}
\hline
\textbf{Notation} & \textbf{Description}              & \textbf{Definition} \\ \hline
$T_{FI}$                                                                         & Fake Instruction         & $NOP$      \\ \hline
$T_{FJ}$                                                                         & Forced JMP               & $(C( JMP, \$label, s_1 , \$label:, s_1 ))$          \\ \hline
$T_{UB}$                                                                         & Untouchable Block        & $(C( JMP, \$label, s_k^+ , \$label: ))$          \\ \hline
$T_{CZJ}$                                                                        & Conditional Zero JMP     & $(C( JZ, \$label, s_1 , \$label:, s_1 ))$          \\ \hline
$T_{CNZJ}$                                                                       & Conditional Non Zero JMP & $(C( JNZ, \$label, s_1 , \$label:, s_1 ))$          \\ \hline
$RE(p,\hat{p})$                                                                         & \begin{tabular}[c]{@{}l@{}}Crossover between 2 parents \\ where $T_{CBI}$ represents \\ code block interchange\end{tabular}   & \begin{tabular}[c]{@{}l@{}}$T\_{CBI}$\{($S^+,LOP^+,COP^+)$,\\      ($\hat{S}^+,\hat{LOP}^+, \hat{COP}^+), C_i\}$\end{tabular}       \\ \hline
\end{tabular}
}
\end{table}

\par
As shown in Figure \ref{fig:pivotPoint}, the pivot point is randomly chosen to be any point beyond non-overlapping block of statements ($s_i$). By virtue of this pivot point, all transformation functions discussed so far (including $T_{CBI}$) can be applied both above and below the pivot point. Each shaded block in Figure \ref{fig:pivotPoint} shows the space available for application of transformation operations as constrained by the pivot point. Since all the transformation functions described so far can be applied either within a block or between blocks, there remains ample opportunity  for these transformation functions to introduce arbitrary complexity in the code yet retaining its capability for execution. Without loss of generality, the pivot point can be considered to be in the middle of the source assembly code thus facilitating equal opportunity to the blocks above and below for arbitrary transformations. It is worth observing that $T_{CBI}$ is a potential candidate transformation function for realizing single-point and multi-point crossover operations (with the latter requiring multiple pivot points).

\subsection{Malware Antigen Generating Evolutionary algorithm (MAGE)}

The Malware Antigen Generating Evolutionary algorithm (MAGE) has been designed based on the framework and the overview of MAGE is shown in algorithm \ref{Proposed EA}. Therefore, MAGE also seamlessly merges into the form of a traditional EA as shown previously by algorithm \ref{Algo1:StructureOfEA}. Each chromosome generated by MAGE, by virtue of the transformation functions, is a potential variant. This implies that MAGE is able to evolve $P \times G$ variants (P is the population and G is the generation) virus variants, making MAGE a generative algorithm. Additionally, following the novelty search ideals, MAGE also identifies the novel chromosomes in each generation and collates them in the form of a unique dataset. While every chromosome is a potential variant, this dataset of unique variants can serve as the antigens using which the antivirus systems can update their signature database. The intention behind the automated malware generation using MAGE is not just to generate virus variants evading AV scanners but to generate variants as diverse as possible in terms of the assembly code structure. 

\begin{algorithm}
  \caption{MAGE as a framework}
  \label{Proposed EA}
  \begin{algorithmic}[1]
    \Procedure{Evolutionary Algo}{$ \zeta$}
      \State Generate Initial Population using transformations $T_i \in \{\psi,\tau,\sigma\}$ with a probability ${p_m}^i$
      
      \While{(EXE Generated) $| |$ (Generations limit yet to be reached)}
        \State  Compute fitness $\xi=  \sqrt{\sum_{i=1}^{P} (\overrightarrow{SI_i},J_i)^2}$ for every candidate $i$ in current population \Comment{EVALUATE}
        \While{maximum population size limit not reached}
        
        \State \texttt{Select two parents using the tournament selection strategy} \Comment{SELECTION}
        \State \texttt{Perform crossover} \Comment{RECOMBINE}
        \State \texttt{Apply code transformation mutation functions ($i.e. T_i \in \{\psi,\tau,\sigma\}$) each with a probability ${p_m}^i$ on each of the resultant offspring} \Comment{MUTATE}
        \State \texttt{Add each evolved child to the next generation population}
        \State \texttt{Identify the novel individuals from current population and add them to a ``Unique'' variant dataset}
        \EndWhile
      \EndWhile
    \EndProcedure
  \end{algorithmic}
\end{algorithm}

\section{Experiments \& Discussion}
In order to validate the capability of the framework in evolving divergent virus variants, MAGE was seeded with the assembly source of \textit{Intruder} - a virus that can infect .EXE files and jump across directories and even drives \cite{littleblackbook}. The \textit{Intruder} virus attaches itself to the end of any .EXE program and takes control of the program when it first starts. The virus first locates all files (including those from the sub directories) possible to infect and verifies if it can be infected before actually infecting the .EXE file. Since \textit{Intruder} is an extremely infectious virus, the test-bed consisted of an isolated computer, inside which a 32bit Microsoft Windows 7 guest virtual machine was used. The host operating system was a 64bit Linux Mint system. This ensured that the experiments that were conducted did not escape the test environment and spread. The host machine had an Intel© Core™ i5-2400 CPU @ 3.10GHz with 4 CPU cores and 8 GB RAM with a 1TB Hard disk. The guest virtual machine used 3 CPU cores and 4GB of RAM with an execution cap of 80\%. Regular snapshots of the virtual machine was taken to keep track of any changes the system might exhibit and thereby identify possible accidental infections. The guest windows OS also had a copy of the Microsoft Macro Assembler (MASM) which is required to make the virus an executable. Also, since the focus is on evolving divergent malware from a single source, no other obfuscation techniques or packers were employed during the test.\\

\par

 \begin{figure}[htbp]
    \centering
    \begin{tikzpicture}
\begin{axis}[
    width=0.49\textwidth,
    xlabel={Best variant from each Generation},
    ylabel={Similarity With Source},
    xmin=0, xmax=310,
    ymin=0.3, ymax=1.2,
    xtick={0,50,100,150,200,250,300},
    ytick={0.3,0.4,0.5,0.6,0.7,0.8,0.9,1.0},
    legend pos=north east,
    ymajorgrids=true,
    xmajorgrids=true,
    axis y discontinuity=crunch,
    grid style=dashed,
    ylabel near ticks
]
\addplot[
    line width=0.25mm,
    color=blue,
    opacity=0.75
    ]
    coordinates { (0,0.961928934010152) (1,0.966836734693877) (2,0.97680412371134) (3,0.974293059125964) (4,0.97680412371134) (5,0.979328165374677) (6,0.979328165374677) (7,0.974293059125964) (8,0.97680412371134) (9,0.984415584415584) (10,0.979328165374677) (11,0.974293059125964) (12,0.971794871794872) (13,0.97680412371134) (14,0.966836734693877) (15,0.979328165374677) (16,0.974293059125964) (17,0.971794871794872) (18,0.979328165374677) (19,0.984415584415584) (20,0.981865284974093) (21,0.97680412371134) (22,0.971794871794872) (23,0.981865284974093) (24,0.979328165374677) (25,0.986979166666667) (26,0.981865284974093) (27,0.971794871794872) (28,0.971794871794872) (29,0.964376590330789) (30,0.971794871794872) (31,0.971794871794872) (32,0.964376590330789) (33,0.974293059125964) (34,0.979328165374677) (35,0.97680412371134) (36,0.981865284974093) (37,0.97680412371134) (38,0.966836734693877) (39,0.969309462915601) (40,0.964376590330789) (41,0.97680412371134) (42,0.974293059125964) (43,0.974293059125964) (44,0.969309462915601) (45,0.974293059125964) (46,0.974293059125964) (47,0.974293059125964) (48,0.97680412371134) (49,0.971794871794872) (50,0.969309462915601) (51,0.966836734693877) (52,0.966836734693877) (53,0.966836734693877) (54,0.964376590330789) (55,0.961928934010152) (56,0.964376590330789) (57,0.984415584415584) (58,0.959493670886076) (59,0.969309462915601) (60,0.97680412371134) (61,0.969309462915601) (62,0.974293059125964) (63,0.97680412371134) (64,0.971794871794872) (65,0.984415584415584) (66,0.981865284974093) (67,0.979328165374677) (68,0.984415584415584) (69,0.979328165374677) (70,0.979328165374677) (71,0.981865284974093) (72,0.971794871794872) (73,0.979328165374677) (74,0.974293059125964) (75,0.981865284974093) (76,0.97680412371134) (77,0.979328165374677) (78,0.984415584415584) (79,0.97680412371134) (80,0.974293059125964) (81,0.979328165374677) (82,0.979328165374677) (83,0.971794871794872) (84,0.974293059125964) (85,0.971794871794872) (86,0.969309462915601) (87,0.981865284974093) (88,0.984415584415584) (89,0.979328165374677) (90,0.981865284974093) (91,0.979328165374677) (92,0.97680412371134) (93,0.981865284974093) (94,0.971794871794872) (95,0.984415584415584) (96,0.981865284974093) (97,0.97680412371134) (98,0.97680412371134) (99,0.969309462915601) (100,0.974293059125964) (101,0.97680412371134) (102,0.97680412371134) (103,0.971794871794872) (104,0.974293059125964) (105,0.969309462915601) (106,0.971794871794872) (107,0.971794871794872) (108,0.981865284974093) (109,0.974293059125964) (110,0.974293059125964) (111,0.979328165374677) (112,0.971794871794872) (113,0.984415584415584) (114,0.969309462915601) (115,0.981865284974093) (116,0.97680412371134) (117,0.974293059125964) (118,0.971794871794872) (119,0.971794871794872) (120,0.969309462915601) (121,0.97680412371134) (122,0.974293059125964) (123,0.974293059125964) (124,0.986979166666667) (125,0.974293059125964) (126,0.981865284974093) (127,0.981865284974093) (128,0.974293059125964) (129,0.979328165374677) (130,0.974293059125964) (131,0.969309462915601) (132,0.961928934010152) (133,0.981865284974093) (134,0.981865284974093) (135,0.984415584415584) (136,0.984415584415584) (137,0.981865284974093) (138,0.971794871794872) (139,0.971794871794872) (140,0.984415584415584) (141,0.986979166666667) (142,0.984415584415584) (143,0.984415584415584) (144,0.971794871794872) (145,0.971794871794872) (146,0.981865284974093) (147,0.974293059125964) (148,0.981865284974093) (149,0.979328165374677) (150,0.984415584415584) (151,0.981865284974093) (152,0.97680412371134) (153,0.981865284974093) (154,0.971794871794872) (155,0.971794871794872) (156,0.97680412371134) (157,0.981865284974093) (158,0.971794871794872) (159,0.974293059125964) (160,0.97680412371134) (161,0.971794871794872) (162,0.971794871794872) (163,0.969309462915601) (164,0.969309462915601) (165,0.97680412371134) (166,0.979328165374677) (167,0.97680412371134) (168,0.984415584415584) (169,0.97680412371134) (170,0.97680412371134) (171,0.969309462915601) (172,0.974293059125964) (173,0.969309462915601) (174,0.974293059125964) (175,0.979328165374677) (176,0.979328165374677) (177,0.979328165374677) (178,0.984415584415584) (179,0.97680412371134) (180,0.981865284974093) (181,0.981865284974093) (182,0.981865284974093) (183,0.981865284974093) (184,0.981865284974093) (185,0.984415584415584) (186,0.981865284974093) (187,0.971794871794872) (188,0.981865284974093) (189,0.974293059125964) (190,0.981865284974093) (191,0.981865284974093) (192,0.979328165374677) (193,0.981865284974093) (194,0.979328165374677) (195,0.981865284974093) (196,0.984415584415584) (197,0.986979166666667) (198,0.984415584415584) (199,0.984415584415584) (200,0.97680412371134) (201,0.974293059125964) (202,0.97680412371134) (203,0.979328165374677) (204,0.984415584415584) (205,0.981865284974093) (206,0.971794871794872) (207,0.971794871794872) (208,0.97680412371134) (209,0.97680412371134) (210,0.974293059125964) (211,0.971794871794872) (212,0.97680412371134) (213,0.969309462915601) (214,0.974293059125964) (215,0.969309462915601) (216,0.966836734693877) (217,0.971794871794872) (218,0.974293059125964) (219,0.964376590330789) (220,0.964376590330789) (221,0.969309462915601) (222,0.97680412371134) (223,0.971794871794872) (224,0.984415584415584) (225,0.971794871794872) (226,0.981865284974093) (227,0.984415584415584) (228,0.992146596858639) (229,0.984415584415584) (230,0.97680412371134) (231,0.981865284974093) (232,0.979328165374677) (233,0.974293059125964) (234,0.971794871794872) (235,0.971794871794872) (236,0.969309462915601) (237,0.97680412371134) (238,0.981865284974093) (239,0.981865284974093) (240,0.981865284974093) (241,0.984415584415584) (242,0.979328165374677) (243,0.979328165374677) (244,0.974293059125964) (245,0.961928934010152) (246,0.974293059125964) (247,0.974293059125964) (248,0.981865284974093) (249,0.986979166666667) (250,0.979328165374677) (251,0.981865284974093) (252,0.97680412371134) (253,0.981865284974093) (254,0.979328165374677) (255,0.974293059125964) (256,0.979328165374677) (257,0.984415584415584) (258,0.97680412371134) (259,0.979328165374677) (260,0.989556135770235) (261,0.97680412371134) (262,0.97680412371134) (263,0.97680412371134) (264,0.981865284974093) (265,0.979328165374677) (266,0.984415584415584) (267,0.979328165374677) (268,0.981865284974093) (269,0.984415584415584) (270,0.984415584415584) (271,0.984415584415584) (272,0.981865284974093) (273,0.979328165374677) (274,0.979328165374677) (275,0.986979166666667) (276,0.989556135770235) (277,0.97680412371134) (278,0.979328165374677) (279,0.984415584415584) (280,0.974293059125964) (281,0.981865284974093) (282,0.984415584415584) (283,0.984415584415584) (284,0.974293059125964) (285,0.979328165374677) (286,0.97680412371134) (287,0.986979166666667) (288,0.979328165374677) (289,0.974293059125964) (290,0.981865284974093) (291,0.974293059125964) (292,0.971794871794872) (293,0.969309462915601) (294,0.974293059125964) (295,0.969309462915601) (296,0.971794871794872) (297,0.961928934010152) (298,0.971794871794872) (299,0.969309462915601) (300,0.969309462915601)   };
     \addplot[
    line width=0.35mm,
    color=red,
    opacity=0.75
    ]
    coordinates { (0,0.959493670886076) (1,0.9475) (2,0.938118811881188) (3,0.926650366748166) (4,0.931203931203931) (5,0.926650366748166) (6,0.902380952380952) (7,0.931203931203931) (8,0.911057692307692) (9,0.902380952380952) (10,0.926650366748166) (11,0.902380952380952) (12,0.926650366748166) (13,0.906698564593301) (14,0.908872901678657) (15,0.915458937198068) (16,0.906698564593301) (17,0.889671361502347) (18,0.904534606205251) (19,0.904534606205251) (20,0.881395348837209) (21,0.887587822014051) (22,0.867276887871853) (23,0.869266055045872) (24,0.863325740318907) (25,0.853603603603604) (26,0.845982142857143) (27,0.825708061002178) (28,0.867276887871853) (29,0.804670912951168) (30,0.831140350877193) (31,0.895981087470449) (32,0.845982142857143) (33,0.834801762114537) (34,0.827510917030568) (35,0.82034632034632) (36,0.851685393258427) (37,0.838495575221239) (38,0.836644591611479) (39,0.823913043478261) (40,0.816810344827586) (41,0.811563169164882) (42,0.825708061002178) (43,0.823913043478261) (44,0.801268498942918) (45,0.801268498942918) (46,0.81505376344086) (47,0.792887029288703) (48,0.808102345415778) (49,0.791231732776618) (50,0.811563169164882) (51,0.792887029288703) (52,0.784679089026915) (53,0.808102345415778) (54,0.792887029288703) (55,0.781443298969072) (56,0.797894736842105) (57,0.764112903225807) (58,0.753479125248509) (59,0.797894736842105) (60,0.771894093686354) (61,0.768762677484787) (62,0.762575452716298) (63,0.764112903225807) (64,0.764112903225807) (65,0.794549266247379) (66,0.737354085603113) (67,0.792887029288703) (68,0.784679089026915) (69,0.733075435203095) (70,0.775051124744376) (71,0.730250481695568) (72,0.730250481695568) (73,0.764112903225807) (74,0.731660231660231) (75,0.754980079681275) (76,0.731660231660231) (77,0.713747645951036) (78,0.744597249508841) (79,0.756487025948104) (80,0.721904761904762) (81,0.746062992125984) (82,0.719165085388994) (83,0.715094339622641) (84,0.756487025948104) (85,0.741682974559687) (86,0.744597249508841) (87,0.730250481695568) (88,0.735922330097087) (89,0.719165085388994) (90,0.735922330097087) (91,0.707089552238806) (92,0.720532319391635) (93,0.705772811918063) (94,0.747534516765286) (95,0.700554528650647) (96,0.728846153846154) (97,0.735922330097087) (98,0.71780303030303) (99,0.731660231660231) (100,0.724665391969407) (101,0.721904761904762) (102,0.690346083788707) (103,0.740234375) (104,0.719165085388994) (105,0.692870201096892) (106,0.711069418386492) (107,0.703153988868275) (108,0.711069418386492) (109,0.708411214953271) (110,0.703153988868275) (111,0.694139194139194) (112,0.695412844036697) (113,0.697974217311234) (114,0.691605839416058) (115,0.691605839416058) (116,0.67921146953405) (117,0.691605839416058) (118,0.681654676258993) (119,0.684115523465704) (120,0.673179396092362) (121,0.676785714285714) (122,0.676785714285714) (123,0.671985815602837) (124,0.674377224199288) (125,0.674377224199288) (126,0.677996422182469) (127,0.677996422182469) (128,0.666080843585237) (129,0.661431064572426) (130,0.666080843585237) (131,0.669611307420495) (132,0.657986111111111) (133,0.660278745644599) (134,0.661431064572426) (135,0.65684575389948) (136,0.668430335097002) (137,0.644557823129252) (138,0.640202702702703) (139,0.666080843585237) (140,0.636974789915966) (141,0.642372881355932) (142,0.642372881355932) (143,0.651202749140893) (144,0.647863247863248) (145,0.664912280701754) (146,0.636974789915966) (147,0.657986111111111) (148,0.632721202003339) (149,0.634840871021776) (150,0.653448275862069) (151,0.630615640599002) (152,0.630615640599002) (153,0.655709342560554) (154,0.626446280991735) (155,0.653448275862069) (156,0.654576856649395) (157,0.630615640599002) (158,0.647863247863248) (159,0.624382207578254) (160,0.614262560777958) (161,0.648972602739726) (162,0.617263843648209) (163,0.614262560777958) (164,0.644557823129252) (165,0.609324758842444) (166,0.646757679180887) (167,0.609324758842444) (168,0.641285956006768) (169,0.607371794871795) (170,0.642372881355932) (171,0.612277867528271) (172,0.609324758842444) (173,0.600633914421553) (174,0.636974789915966) (175,0.608346709470305) (176,0.603503184713376) (177,0.623355263157895) (178,0.632721202003339) (179,0.6064) (180,0.607371794871795) (181,0.628524046434494) (182,0.620294599018003) (183,0.607371794871795) (184,0.598736176935229) (185,0.621311475409836) (186,0.601587301587302) (187,0.625412541254125) (188,0.595911949685535) (189,0.626446280991735) (190,0.598736176935229) (191,0.602543720190779) (192,0.597791798107255) (193,0.596850393700787) (194,0.622331691297209) (195,0.598736176935229) (196,0.622331691297209) (197,0.593114241001565) (198,0.617263843648209) (199,0.596850393700787) (200,0.618270799347471) (201,0.618270799347471) (202,0.594976452119309) (203,0.591263650546022) (204,0.599683544303797) (205,0.600633914421553) (206,0.613268608414239) (207,0.609324758842444) (208,0.609324758842444) (209,0.605431309904153) (210,0.601587301587302) (211,0.601587301587302) (212,0.600633914421553) (213,0.594976452119309) (214,0.593114241001565) (215,0.588509316770186) (216,0.594043887147335) (217,0.586687306501548) (218,0.582181259600614) (219,0.594043887147335) (220,0.575987841945289) (221,0.590342679127726) (222,0.572507552870091) (223,0.569069069069069) (224,0.580398162327718) (225,0.562314540059347) (226,0.558997050147493) (227,0.575113808801214) (228,0.573373676248109) (229,0.565671641791045) (230,0.575113808801214) (231,0.562314540059347) (232,0.575987841945289) (233,0.561481481481482) (234,0.558173784977909) (235,0.572507552870091) (236,0.545323741007194) (237,0.565671641791045) (238,0.540656205420827) (239,0.53988603988604) (240,0.540656205420827) (241,0.554904831625183) (242,0.551673944687045) (243,0.540656205420827) (244,0.540656205420827) (245,0.53758865248227) (246,0.536827195467422) (247,0.54297994269341) (248,0.536827195467422) (249,0.546897546897547) (250,0.533802816901408) (251,0.54297994269341) (252,0.541428571428571) (253,0.536827195467422) (254,0.533802816901408) (255,0.525658807212205) (256,0.532303370786517) (257,0.519890260631001) (258,0.525658807212205) (259,0.524204702627939) (260,0.516348773841962) (261,0.533052039381153) (262,0.524204702627939) (263,0.526388888888889) (264,0.519890260631001) (265,0.520604395604396) (266,0.522038567493113) (267,0.519890260631001) (268,0.527855153203343) (269,0.524930747922438) (270,0.514945652173913) (271,0.518467852257182) (272,0.522038567493113) (273,0.514945652173913) (274,0.523480662983425) (275,0.519178082191781) (276,0.522038567493113) (277,0.520604395604396) (278,0.502652519893899) (279,0.51775956284153) (280,0.506008010680908) (281,0.50332005312085) (282,0.509408602150538) (283,0.509408602150538) (284,0.512162162162162) (285,0.519178082191781) (286,0.518467852257182) (287,0.48840206185567) (288,0.51775956284153) (289,0.48840206185567) (290,0.512855209742896) (291,0.489032258064516) (292,0.485897435897436) (293,0.50668449197861) (294,0.492207792207792) (295,0.5) (296,0.487146529562982) (297,0.493489583333333) (298,0.496073298429319) (299,0.49737532808399) (300,0.483418367346939) };
  \legend{Fitness type $\alpha$, Fitness type $\beta$}
    
\end{axis}
\end{tikzpicture}
  \caption{Similarity values of Initial and Final population using $\alpha$ and $\beta$}
    \label{fig:absolouteVsNovelty}
\end{figure}
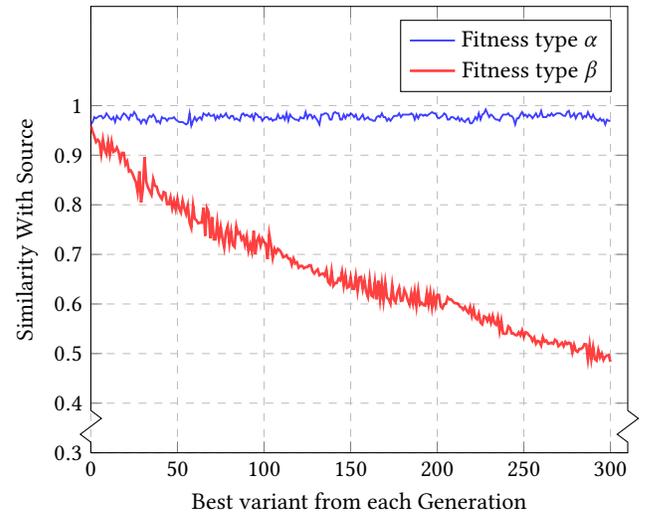
\par

\par

The purpose of the experiments was to verify if a novelty supported intra-population similarity based approach is capable of generating malware variants of greater diversity than a simple similarity-only based approach. Therefore, the simulation experiments involved evaluating the diversity, in terms of code structures, the result of applying MAGE with two different quality indicators (fitness functions) $\alpha$ and $\beta$. The indicator $\alpha$ solely depended on Jaccard similarity as a metric for evaluation, while the fitness function $\beta$ exploited the Jaccard similarity index based intra-population similarity. It is also worth mentioning that the focus of the work is on a framework for evolving divergent variants rather than bench-marking. In order to ensure fairness during comparison, the same random seed was set and used for the experiments and only the Jaccard similarity best fit individuals from each generation were compared with the source virus. The population size has been set at a random value of 20 and the EA has been run for 300 generations. The transformation functions $T_{FI}$, $T_{FJ}$, $T_{UB}$, $T_{CZJ}$ and $T_{CNZJ}$ are employed to mutate assembly code structure and $T_{CBI}$ is employed for crossover operation. The results (figure \ref{fig:absolouteVsNovelty}) show that a novelty supported intra-population similarity based fitness function is able to induce more variations within the population.\\

\par
A statistical analysis using the Mann-Whitney U test was also performed on the similarity value of the variants evolved in both the initial and final populations. The Mann–Whitney U test is a popular non-parametric hypothesis test that verifies a null hypothesis ($H_0$) and a research hypothesis ($H_1$). In the current analysis, the null hypothesis ($H_0$) states that the initial and final populations are the same. Conversely, in this context, the research hypothesis ($H_1$) states that the initial and final populations are different. The test calculates the value of $U$ which ranges from 0 to $n_1 \times n_2$, where $n_1$ and $n_2$ are the sizes of each population. Based on the probability ($p$) that the results appear by chance, a high value of U accepts the null hypothesis ($H_0$) and rejects the research hypothesis ($H_1$). Conversely, a low value of U rejects the null hypothesis ($H_0$) and accepts the research hypothesis ($H_1$), also based on the probability ($p$) of the results appearing by chance. In the case of most biological analyses, it is usually admissible to use $p < 0.01$ as the threshold for acceptance for the Mann-Whitney U test.

\begin{table}[htbp]
\centering
\caption{Similarity Values of Initial population vs Final population}
\label{Table:Similarity_InitVsFinal}
\scalebox{0.70}{
\begin{tabular}{|l|ll|ll|}
\hline
   & \multicolumn{2}{c|}{\textbf{Similarity with Source $(\alpha)$}}                         & \multicolumn{2}{c|}{\textbf{Novelty $(\beta)$}}                                        \\ \hline
   & \multicolumn{1}{l|}{\textit{Initial Population}} & \textit{Final Population} & \multicolumn{1}{l|}{\textit{Initial Population}} & \textit{Final Population} \\ \hline
1  & \multicolumn{1}{l|}{0.9844155844}                & 0.9973684211              & \multicolumn{1}{l|}{0.981865285}                 & 0.4834183673              \\ \hline
2  & \multicolumn{1}{l|}{0.9717948718}                & 0.9793281654              & \multicolumn{1}{l|}{0.981865285}                 & 0.4922077922              \\ \hline
3  & \multicolumn{1}{l|}{0.9973684211}                & 0.9921465969              & \multicolumn{1}{l|}{0.9742930591}                & 0.500660502               \\ \hline
4  & \multicolumn{1}{l|}{0.9844155844}                & 0.9973684211              & \multicolumn{1}{l|}{0.9973684211}                & 0.4947780679              \\ \hline
5  & \multicolumn{1}{l|}{0.981865285}                 & 0.9844155844              & \multicolumn{1}{l|}{0.961928934}                 & 0.4884020619              \\ \hline
6  & \multicolumn{1}{l|}{0.9717948718}                & 0.9973684211              & \multicolumn{1}{l|}{0.9793281654}                & 0.4896640827              \\ \hline
7  & \multicolumn{1}{l|}{0.9947506562}                & 0.9973684211              & \multicolumn{1}{l|}{0.9947506562}                & 0.5060080107              \\ \hline
8  & \multicolumn{1}{l|}{0.9844155844}                & 0.9768041237              & \multicolumn{1}{l|}{1}                           & 0.4834183673              \\ \hline
9  & \multicolumn{1}{l|}{0.961928934}                 & 0.9717948718              & \multicolumn{1}{l|}{0.9869791667}                & 0.5039893617              \\ \hline
10 & \multicolumn{1}{l|}{1}                           & 0.9973684211              & \multicolumn{1}{l|}{0.9793281654}                & 0.506684492               \\ \hline
11 & \multicolumn{1}{l|}{0.9844155844}                & 0.9895561358              & \multicolumn{1}{l|}{0.9768041237}                & 0.4973753281              \\ \hline
12 & \multicolumn{1}{l|}{1}                           & 0.981865285               & \multicolumn{1}{l|}{0.981865285}                 & 0.4980289093              \\ \hline
13 & \multicolumn{1}{l|}{0.981865285}                 & 0.981865285               & \multicolumn{1}{l|}{0.9895561358}                & 0.49672346                \\ \hline
14 & \multicolumn{1}{l|}{0.981865285}                 & 0.9844155844              & \multicolumn{1}{l|}{0.981865285}                 & 0.5013227513              \\ \hline
15 & \multicolumn{1}{l|}{1}                           & 0.9947506562              & \multicolumn{1}{l|}{0.981865285}                 & 0.4947780679              \\ \hline
16 & \multicolumn{1}{l|}{0.9895561358}                & 0.9973684211              & \multicolumn{1}{l|}{0.961928934}                 & 0.4947780679              \\ \hline
17 & \multicolumn{1}{l|}{0.9643765903}                & 0.9768041237              & \multicolumn{1}{l|}{0.9594936709}                & 0.4922077922              \\ \hline
18 & \multicolumn{1}{l|}{0.9768041237}                & 0.9793281654              & \multicolumn{1}{l|}{0.9844155844}                & 0.5039893617              \\ \hline
19 & \multicolumn{1}{l|}{0.9693094629}                & 0.9947506562              & \multicolumn{1}{l|}{0.9869791667}                & 0.5039893617              \\ \hline
20 & \multicolumn{1}{l|}{0.9844155844}                & 0.9895561358              & \multicolumn{1}{l|}{0.9844155844}                & 0.506684492               \\ \hline
\end{tabular}
}
\end{table}

On performing the Mann-Whitney U Test on the initial and final populations (shown in Table \ref{Table:Similarity_InitVsFinal}) resulting from using $\alpha$ as the fitness metric in MAGE, the $U,p$ and $z score$ values were calculated. Both $U = 161$ and $z score = -1.04143$ fall within the acceptable range (where $128.065$ to $271.935$ is the region of acceptance for U and between $-1.96$ and $1.96$ is the region of acceptance for z score), and the value of $p$ was $0.29834$, which is not within the permissible threshold limit of $0.01$. This implied that the hypothesis $H_0$ cannot be conclusively rejected and so the two populations are considered to be similar. On the other hand, when the Mann-Whitney U Test was performed on the initial and final populations resulting from using $\beta$ as the fitness metric in MAGE, the difference between the randomly selected values of the initial and the final populations was large enough to be statistically significant. The value of $U$ was $400$ and $z score$ was $5.4054$ (both outside the region of acceptance, namely from 127.6622 to 272.3378 for U and from -1.96 to 1.96 for z score) and $p$ was $6.467 \times 10^{-8}$ which is definitely less than the threshold of $0.01$. From this it is clear that the hypothesis $H_0$ can be conclusively rejected and the research hypothesis $H_1$ (which states that the two populations are different) can be accepted. To further illustrate that the evolution was indeed encouraged by the cascading effect of the fitness function, the Mann-Whitney U Test was also applied on the initial populations of both distributions. The results ($U=223, z score = 0.6133,p=0.5397$) show that null hypothesis is true and both the populations are same. Since initial populations of both the distributions are level, it can be concluded that the diversity in the populations was more pronounced the during evolution, when using a novelty search based fitness function ($\beta$). This further supports the hypothesis that a novelty search based approach ($\beta$) is capable of generating malware variants of greater diversity than a simple similarity based ($\alpha$) approach. \\

\par

 \begin{figure}[htbp]
    \centering
    \begin{tikzpicture}
\begin{axis}[
    width=0.49\textwidth,
    xlabel={Best variant from each Generation},
    ylabel={No. of AV scanners detecting $\zeta'$},
    xmin=0, xmax=310,
    ymin=-1, ymax=22,
    xtick={0,50,100,150,200,250,300},
    ytick={0,2,4,6,8,10,12,14,16,18,20},
    legend pos=north east,
    ymajorgrids=true,
    xmajorgrids=true,
    grid style=dashed,
    ylabel near ticks
]
\addplot[
    line width=0.25mm,
    color=blue,
    opacity=0.75
    ]
    coordinates { (0,6) (1,7) (2,8) (3,8) (10,6) (11,5) (12,6) (13,6) (14,6) (15,8) (16,5) (17,8) (18,5) (19,8) (20,7) (21,7) (22,6) (23,7) (24,7) (25,3) (26,6) (27,7) (28,4) (29,6) (30,4) (31,4) (32,4) (33,4) (34,4) (35,4) (36,4) (37,3) (38,4) (39,4) (40,4) (100,4) (101,0) (102,4) (103,0) (104,3) (105,4) (106,5) (107,5) (108,4) (109,4) (110,4) (111,3) (112,3) (113,0) (114,4) (115,3) (116,4) (117,4) (118,4) (119,4) (120,4) (121,1) (122,0) (123,4) (124,3) (125,0) (126,4) (127,3) (128,4) (129,3) (130,3) (131,0) (132,3) (133,4) (134,3) (135,3) (136,4) (137,0) (138,4) (139,3) (140,0) (141,4) (142,0) (143,4) (144,0) (145,3) (146,4) (147,4) (148,3) (149,3) (150,4) (151,4) (152,3) (153,0) (154,4) (155,0) (156,4) (157,3) (158,4) (159,4) (160,0) (161,4) (162,0) (163,4) (164,3) (165,4) (166,20) (167,3) (168,3) (169,0) (170,3) (171,0) (172,0) (173,3) (174,3) (175,0) (176,3) (177,3) (178,0) (179,0) (180,3) (181,0) (182,0) (183,0) (184,3) (185,2) (186,0) (187,0) (188,0) (189,3) (190,0) (191,0) (192,0) (193,0) (194,0) (195,3) (196,3) (197,0) (198,0) (199,0) (200,3) (201,0) (202,3) (203,0) (204,3) (205,3) (206,3) (207,0) (208,3) (209,0) (210,0) (211,3) (212,0) (213,3) (214,0) (215,0) (216,0) (217,3) (218,0) (219,0) (220,3) (221,2) (222,0) (223,0) (224,3) (225,0) (226,3) (227,0) (228,2) (229,0) (230,0) (231,2) (232,3) (233,0) (234,0) (235,3) (236,3) (237,0) (238,3) (239,3) (240,3) (241,0) (242,0) (243,3) (244,0) (245,3) (246,3) (247,0) (248,3) (249,3) (250,3) (251,0) (252,0) (253,3) (254,0) (255,3) (256,0) (257,0) (258,0) (259,0) (260,0) (261,0) (262,0) (263,0) (264,0) (265,0) (266,0) (267,0) (268,0) (269,0) (270,0) (271,0) (272,0) (273,0) (274,0) (275,0) (276,0) (277,0) (278,0) (279,0) (280,0) (281,0) (282,0) (283,0) (284,0) (285,0) (286,0) (287,0) (288,0) (289,0) (290,0) (291,0) (292,0) (293,0) (294,0) (295,0) (296,0) (297,0) (298,0) (299,0) (300,0) };
    
    
\end{axis}
\end{tikzpicture}
  \caption{Evasion capability of MAGE: Number of antivirus scanners detecting the generated variants}
    \label{fig:evasionCapability}
\end{figure}
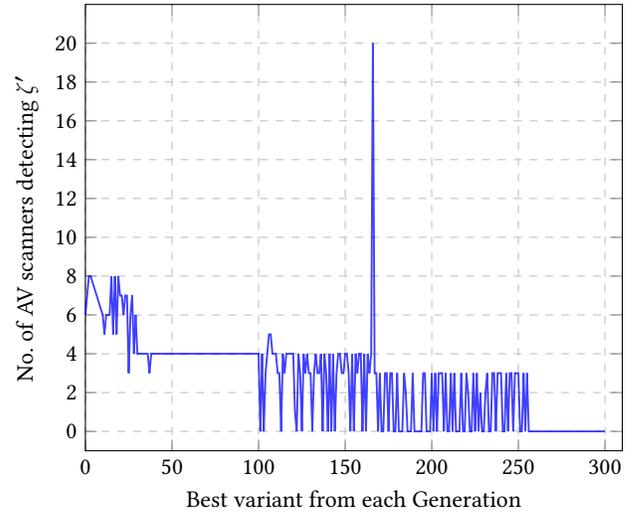

\par
At this juncture, it is worth mentioning that, since the aim of the EA was to evolve divergent variants that are capable of evading antivirus scanners while not altering the maliciousness of the resultant variant, the evasion capability of the virus and its variants was also checked using VirusTotal \cite{virustotal} which used over 60 antivirus scanners to scan each executable. The VirusTotal results showed that 20 popular antivirus scanners were able to identify the source \textit{Intruder} virus. However, it was observed that by the $250^{th}$ generation, almost all the antivirus scanners used by VirusTotal (over 98\%) were evaded by variants evolved. By way of example, figure \ref{fig:evasionCapability} displays evasion capability of variants evolved by MAGE when using $\beta$ as the quality indicator. This observation was expected as it is a well known fact that even minor modifications in code is sufficient to evade most antivirus scanners. It is also worth noting that, by virtue of the transformation functions ($T$) and well defined recombination operator ($R$), MAGE is capable of evolving valid virus executables for over 600 generations i.e. $600 \times 20 = 12,000$ divergent virus variables successfully. This is because ``bloat'' is a common factor in EAs \cite{banzhaf1998genetic} and the Microsoft Macro Assembler (MASM) required that the files remained under the 64KB limit.  \\
\par

The above simulation experiments have demonstrated that a novelty supported intra-population similarity based approach is capable of generating diverse variants of a given malware. Subsequent Mann-Whitney U test based statistical analysis on the similarity value of the variants evolved revealed that the diversity in the populations was more pronounced during the evolution, when using a novelty search based fitness function ($\beta$). The validity of the framework was also demonstrated by the proposed Malware Antigen Generating Evolutionary algorithm
(MAGE) which evolved diverse virus variants that evaded detection by over 98\% of all antivirus scanners in VirusTotal. \\
\par

\section{Conclusion}
This work discussed a generic assembly source code based framework that facilitates an evolutionary algorithm to generate diverse and potential variants of an input malware, while retaining its maliciousness, yet capable of evading antivirus scanners. The generic code transformation functions $T \subseteq \{\psi, \tau,\sigma\}$ based on which five transformation function instances namely Fake Instruction ($T_{FI}$), Forced JMP ($T_{FJ}$), Untouchable Block ($T_{UB}$), Conditional JMP ($T_{CZJ}$) and Conditional JMP ($T_{CNZJ}$), were also proposed and defined as mutation operators. The code block interchange transformation function ($T_{CBI}$) was utilized in the design of the crossover operator to enable seamless recombination resulting in a valid executable.  \\
\par
The validity of the framework was also demonstrated by the proposed Malware Antigen Generating Evolutionary algorithm (MAGE) which utilized a novelty search supported intra-population based fitness function to evolve diverse variants of a source malware. The simulation experiments were performed using \textit{Intruder} - a virus that attaches itself to the end of any .EXE program. The versatility, efficacy and flexibility of the MAGE was also demonstrated by utilizing two different quality indicators (fitness functions) $\alpha$ and $\beta$, to evolve valid variants of the \textit{Intruder} virus. VirusTotal was utilized to observe the evasion capability of the evolved \textit{Intruder} virus variants. Results show that almost all the antivirus scanners were evaded by the evolved variants after 250 generations of evolution. Statistical analysis on the initial and final populations further revealed that the novelty supported intra-population based fitness function ($\beta$) was able to direct MAGE to evolve divergent malware variants of the given source malware.\\
\par
Since every candidate evolved by MAGE is a potential variant, the entire collection of variants generated could serve as the malware variant dataset. This dataset of antigens could then be presented to the malware analysis engines to improve their malware detection algorithms. Thus, in conclusion, the framework and MAGE could serve as a flexible platform for researchers and practitioners to develop novel applications under the aegis of automated malware generation for proactive defence.

 \newcommand{\noop}[1]{}

\end{document}